\documentclass[10pt,conference]{IEEEtran}
\usepackage{cite}
\usepackage{amsmath,amssymb,amsfonts}
\usepackage{graphicx}
\usepackage{textcomp}
\usepackage{xcolor}
\usepackage{booktabs}
\usepackage{multirow}
\usepackage{array}
\usepackage{microtype}
\usepackage{placeins}
\usepackage{hyperref}

\graphicspath{{figures/}}

\title{CourseBlueprint: A Structured Pipeline for Adaptive Pedagogical Video Generation Grounded in Course Corpora}
\author{
  Md~Zabirul~Islam,
  Md~Motaleb~Hossen~Manik,
  and~Ge~Wang$^{*}$%
  \thanks{M.~Z.~Islam and M.~M.~H.~Manik are with the Department of Computer
  Science, Rensselaer Polytechnic Institute, Troy, NY 12180, USA (e-mail:
  islamm11@rpi.edu; manikm@rpi.edu).}%
  \thanks{G.~Wang is with the Department of Biomedical Engineering,
  Rensselaer Polytechnic Institute, Troy, NY 12180, USA (e-mail:
  wangg6@rpi.edu).}%
  \thanks{$^{*}$Corresponding author: G.~Wang (e-mail: wangg6@rpi.edu).}%
}

\begin{document}
\maketitle

\begin{abstract}
Generative text-to-video systems can produce visually fluent educational clips, but they rarely encode the \emph{pedagogical content knowledge} (PCK) needed for effective instruction, including prerequisite-aware sequencing, learner-adaptive depth, and sustained cognitive engagement. We present \emph{CourseBlueprint}, a course-grounded pipeline for adaptive pedagogical video generation. Given a topic and learner persona, the system generates a structured teaching blueprint in a single forward pass over an undergraduate biomedical-imaging corpus (BMED~2300; twenty-three lectures, $1{,}116$ slides). Instead of ad-hoc prompt chaining, the pipeline uses typed intermediate representations with validation: a scaffolding module builds a stage-labeled prerequisite concept graph with deterministic cycle removal, an adaptive controller assigns per-concept style specifications, and an engagement generator produces narration following a fixed hook$\to$retrieval$\to$core$\to$analogy$\to$forward contract. A deterministic slide-image override further grounds the rendered video by reusing instructor slides whenever retrieval confidence is high. We also release a reusable benchmark corpus and an evaluation harness combining repeated LLM-judge scoring with regex-grounded objective metrics. In a five-topic ablation, removing the engagement contract reduces the engagement score from~5.00 to~1.20, the adaptive score from~4.80 to~3.40, Flesch readability from~38.0 to~19.8, and analogy and retrieval-prompt counts to near zero. The slide-image override converts a $0/9$ corpus-grounding failure into $9/10$ successful slide matches on the same topic. These results show that pedagogical video quality depends less on surface fluency than on explicit, typed instructional contracts that make scaffolding, adaptation, engagement, and grounding auditable.
\end{abstract}

\section{Introduction}

Educational video is a high-impact medium for teaching complex concepts, but producing a high-quality lecture video remains expensive: a single thirty-minute lecture can require hours of expert planning, slide design, narration, and editing. Recent text-to-video systems can now generate visually fluent clips from short prompts~\cite{sora2024,veo2024,lumiere2024,stablevideodiffusion2023,videopoet2024}. This progress shifts the central question from whether generated videos can look plausible to whether they can teach effectively.

Effective teaching requires more than fluent narration and coherent visuals. It requires \emph{pedagogical content knowledge} (PCK): the ability to sequence concepts by prerequisite structure, adapt depth and vocabulary to the learner, and sustain attention through meaningful engagement~\cite{shulman1986,magnusson1999,mishra2006tpack}. The First AI~Teaching Monster Challenge~\cite{teachingmonster2026} makes this requirement explicit by evaluating generated instructional videos along three PCK dimensions: knowledge scaffolding, adaptive teaching, and cognitive engagement. These dimensions correspond to established principles in instructional design: prerequisite-aware decomposition of concepts~\cite{wood1976,vygotsky1978}, learner-conditioned explanation~\cite{tomlinson1999,koedinger2012knowledge}, and engagement through curiosity gaps, analogies, and retrieval prompts~\cite{loewenstein1994,roediger2011,chi2014icap}.

Current generative-video baselines remain poorly aligned with these requirements. They typically expand a topic into an outline, convert that outline into slide and narration specifications, and render the final audiovisual artifact. This design treats instructional video as generic content generation. As a result, concepts may appear in an arbitrary order, explanation depth may remain fixed across learner profiles, and engagement moves may occur only incidentally. The output can be fluent while still lacking an explicit instructional structure.

We address this gap with \emph{CourseBlueprint}, a retrieval-augmented pipeline for adaptive pedagogical video generation. The key idea is structural: replace free-form prompt chaining with typed instructional contracts. Given a topic and learner persona, the system generates a sequence of validated intermediate representations grounded in a course corpus. A scaffolding module builds a stage-labeled prerequisite concept graph; an adaptive controller assigns per-concept style specifications; and an engagement generator produces narration under a fixed hook$\to$retrieval$\to$core$\to$analogy$\to$forward template. These typed representations make the instructional plan inspectable, enforceable, and ablation-friendly.

The pipeline is grounded in an undergraduate biomedical-imaging course, BMED~2300, containing twenty-three lectures and $1{,}116$ slide-transcript pairs. Retrieval is used not only to ground the narration but also to control the visual layer. When retrieval identifies a high-confidence course slide, a deterministic slide-image override inserts the instructor's original slide directly into the rendered video, bypassing text-to-image generation. This avoids a common failure mode in which generated visuals drift from the source material despite correct textual grounding.

This paper makes four contributions:
\begin{enumerate}
\item We introduce \emph{CourseBlueprint}, a typed PCK pipeline that integrates scaffolding, adaptive teaching, and engagement in a single forward pass. Its intermediate representations---a concept tree, styled tree, enriched script, and course blueprint---make pedagogical structure auditable and support controlled ablations.

\item We develop a deterministic slide-image override that reuses corpus slides when retrieval confidence is high. This mechanism replaces fragile prompt-marker strategies and improves corpus-slide recovery from $0/9$ to $9/10$ on the same topic.

\item We package a reusable benchmark corpus and evaluation harness built from BMED~2300, including twenty-three lectures, $1{,}116$ slides, aligned transcripts, a multi-repetition LLM-judge rubric, and regex-grounded objective metrics.

\item We provide empirical evidence that structured engagement is the load-bearing PCK component in this setting. In a five-topic ablation, removing the engagement contract reduces the engagement score from~5.00 to~1.20, the adaptive score from~4.80 to~3.40, and Flesch readability from~38.0 to~19.8, while reducing analogy and retrieval-prompt counts to near zero.
\end{enumerate}

The remainder of the paper is organized as follows. Section~\ref{sec:related} reviews related work in generative video, intelligent tutoring, retrieval-augmented generation, and structured generation. Section~\ref{sec:overview} presents the system architecture. Section~\ref{sec:method} describes the typed pedagogy modules and audiovisual stack. Sections~\ref{sec:corpus} and~\ref{sec:harness} introduce the benchmark corpus and evaluation methodology. Sections~\ref{sec:experiments} and~\ref{sec:results} report the experimental setup and results. Sections~\ref{sec:disc} and~\ref{sec:limits} discuss interpretation and limitations. Section~\ref{sec:repro} describes reproducibility, and Section~\ref{sec:conc} concludes.

\section{Related Work}\label{sec:related}

\textbf{Generative video foundation models.}
Recent text-to-video systems~\cite{sora2024,veo2024,lumiere2024,stablevideodiffusion2023,videopoet2024} can synthesize visually plausible clips from short prompts, with progress typically measured by visual quality, motion realism, identity consistency, and prompt adherence. These capabilities make instructional-video generation technically feasible, but they do not by themselves address whether the generated artifact teaches. Prompt-only wrappers can produce lecture-shaped outputs, yet they usually delegate instructional structure to a single free-form language-model call. In contrast, our work treats pedagogy as a system-level constraint: we retain the released Teaching Monster starter kit~\cite{teachingmonster2026} for rendering support, but replace its outline and wrapper stages with a typed, course-grounded pedagogy pipeline.

\textbf{Pedagogical AI tutors.}
Intelligent tutoring systems have long modeled aspects of pedagogical content knowledge explicitly~\cite{koedinger1997,vanlehn2006}. More recent language-model tutors~\cite{kasneci2023chatgpt,bailey2025ai,wang2024tutorcopilot,macneil2023code,guo2025one} focus mainly on dialogic settings, where adaptation and engagement can unfold through repeated learner interaction. Asynchronous lecture video is different: the system must precompute the instructional sequence, learner adaptation, and engagement strategy before the learner responds. The AI Teacher Test~\cite{tack2022bea} formalized pedagogical-move evaluation for conversational agents; our work adapts this principle to generated instructional videos by making scaffolding, adaptation, and engagement explicit properties of the video-generation pipeline.

\textbf{Retrieval-augmented and grounded generation.}
Retrieval-augmented generation has become a standard approach for grounding language-model outputs in external sources~\cite{lewis2020,karpukhin2020dpr,asai2024selfrag,zhang2024raft}. Prior work has mainly applied retrieval to question answering, tutoring dialogue, and domain-specific text generation. We use retrieval earlier in the generation process: the course corpus grounds the concept scaffold, the narration, and the visual layer. Slide-granular indexing is central to this design because each retrieved unit links a textual explanation to a specific slide image. The resulting FAISS-based index~\cite{johnson2017faiss} therefore supports both semantic grounding and deterministic slide reuse.

\textbf{Slide and lecture generation from documents.}
Document-to-slide systems~\cite{fu2022doc2ppt,mondal2024slidegen} convert long documents into presentation-style outputs, but the slide deck is usually the endpoint. Our setting requires a complete instructional video, where the visual sequence, spoken narration, learner adaptation, and engagement strategy must remain aligned. We therefore extend beyond document-to-slide generation by adding a typed PCK contract over both the visual and spoken layers.

\textbf{Structured generation and typed contracts.}
Constrained and schema-guided generation methods~\cite{willard2023outlines,poesia2022synchromesh,geng2023grammar} reduce brittleness by forcing model outputs to satisfy formal structure. Our pipeline applies the same idea at the instructional-system level. Each inter-stage payload is a typed object with closed-enumeration fields and validation, so pedagogical requirements are not merely requested in a prompt but enforced as part of the generation contract. This design also makes ablations cleaner: a module can be disabled by replacing its typed object rather than by weakening an instruction.

\textbf{Engagement and multimedia learning.}
Research on online lecture videos shows that production style, pacing, and instructional design affect learner persistence~\cite{guo2014mooc}. Multimedia learning theory further emphasizes segmentation, signaling, and coordinated verbal-visual presentation~\cite{mayer2014mml}. We operationalize these ideas through a structured engagement contract: each generated narration is required to include explicit instructional moves such as a curiosity gap, retrieval prompt, analogy, and forward link. This makes engagement measurable rather than incidental.

\textbf{LLM-as-judge evaluation.}
Rubric-based LLM judges are widely used for evaluating generated text and interactive systems~\cite{zheng2023,liu2023geval,chiang2024arena}, but they can exhibit self-preference when the judge and generator come from the same model family~\cite{panickssery2024selfpref}. Proposed mitigations include calibrated reference judges~\cite{kim2024prometheus2}, evaluator ensembles~\cite{li2023alpacaeval}, and repeated scoring. Our evaluation harness follows this direction by reporting repeated judge scores alongside regex-grounded objective metrics. We also identify cross-model-family judging as an important remaining validation step.

\section{System Overview}\label{sec:overview}

\begin{figure*}[!t]
\centering
\includegraphics[width=0.96\textwidth]{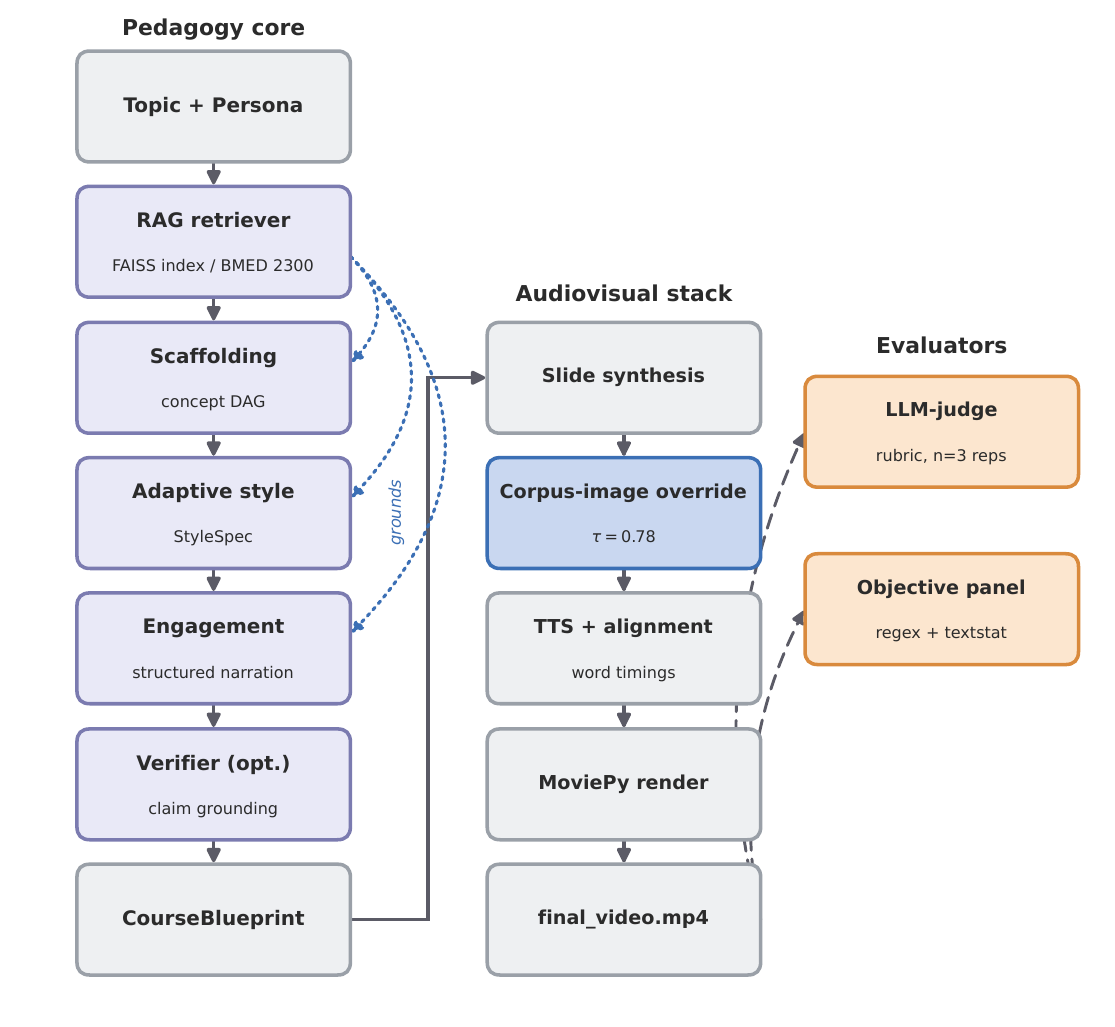}
\caption{System architecture. A topic and learner persona are processed by a
course-grounded pedagogy core consisting of scaffolding, adaptive style,
engagement, and an optional verifier. Each module is grounded by FAISS
retrieval over the BMED~2300 corpus and emits typed intermediate
representations that are assembled into a course blueprint. The blueprint
drives the audiovisual stack, where high-confidence retrieval matches trigger
a deterministic corpus-slide override, audio is generated by text-to-speech,
word timings are recovered by forced alignment, and MoviePy renders the final
video. The generated artefacts are evaluated by repeated LLM-judge scoring and
regex-grounded objective metrics.}
\label{fig:arch}
\end{figure*}

Given a topic string $t$ and a freeform persona string $p$, the system
generates a course blueprint $\mathcal{B}$ and renders it as an instructional
video (Fig.~\ref{fig:arch}). The pipeline has three coordinated components:
a pedagogy core, an audiovisual stack, and an evaluation harness.

\textit{Pedagogy core.}
The pedagogy core converts $(t,p)$ into a typed instructional plan. A FAISS
retriever first grounds the topic in the BMED~2300 course corpus. The
scaffolding module constructs a prerequisite-aware concept structure; the
adaptive style module assigns learner-conditioned style specifications; the
engagement module produces structured narration; and an optional verifier
checks factual grounding. These modules communicate through validated
intermediate representations and jointly produce $\mathcal{B}$. Each module
can be disabled through a Boolean flag, enabling controlled ablations by
replacing its typed output with a default object rather than modifying
free-form prompts. Section~\ref{sec:method} details each module.

\textit{Audiovisual stack.}
The audiovisual stack renders $\mathcal{B}$ into a video. The slide-synthesis
stage uses the blueprint to generate visual layouts, but bypasses image
generation when retrieval provides a high-confidence corpus-slide match. In
those cases, a deterministic slide-hints map inserts the instructor's original
slide image directly into the layout. The narration is synthesized with
text-to-speech, aligned to recover word-level timings, and combined with the
rendered slides using MoviePy. Section~\ref{sec:av} describes the rendering
procedure.

\textit{Evaluation harness.}
The evaluation harness scores both the typed intermediates and the rendered
video. It combines repeated LLM-judge evaluations of scaffolding, adaptation,
and engagement with deterministic objective metrics, including regex-based
counts of engagement moves and textstat-based readability measures. This
pairing allows rubric scores to be checked against measurable properties of
the generated artefact. Section~\ref{sec:harness} describes both evaluators.

\section{Method}\label{sec:method}

This section describes the full generation pipeline introduced in
Section~\ref{sec:overview}. We first define the typed contract that
governs all inter-stage payloads (Section~\ref{sec:schemas}), then
present the retrieval layer, the three core pedagogy modules, the
optional verifier, and the audiovisual stack in pipeline order.

\subsection{Schemas as a Typed Contract}\label{sec:schemas}

\begin{figure*}[!t]
\centering
\includegraphics[width=0.96\textwidth]{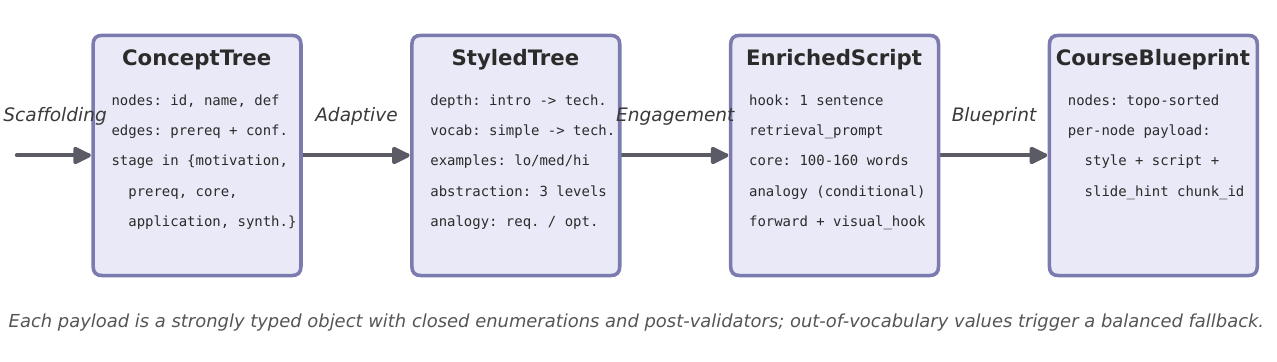}
\caption{Schemas as a typed contract. Each downstream module receives
a strongly typed object and emits another. The closed enumerations
of the style specification and the structured field set of the
enriched script are post-validated; out-of-vocabulary values trigger
a balanced fallback.}
\label{fig:schemas}
\end{figure*}

The pipeline replaces prompt chaining with validated typed
representations. Each module receives a strongly typed object and
emits another object whose fields are constrained by closed
enumerations and post-validators (Fig.~\ref{fig:schemas}). Three
main representations are passed through the pipeline: a concept tree,
a styled tree, and a course blueprint. A nested enriched-script object
stores the per-slide narration.

This contract serves two purposes. First, it makes generation
auditable: scaffolding, adaptation, engagement, and slide grounding
are represented as inspectable fields rather than implicit prompt
effects. Second, it makes ablations meaningful: disabling a module
replaces its typed output with a default object rather than weakening
a free-form instruction. The same structure also allows the objective
evaluator to count engagement moves directly from typed fields.

\subsection{Course-Grounded Retrieval}\label{sec:rag}

Ungrounded language models can replace course-specific explanations
with generic prior knowledge or hallucinated domain details. To
prevent this drift, every pedagogy stage queries a retrieval index
built from the BMED~2300 corpus.

The corpus is chunked at slide granularity. Each chunk contains the
per-slide transcript, lecture and slide indices, a stable chunk
identifier, and the corresponding slide-image path. The resulting
$1{,}116$ chunks are embedded with Google's general-purpose embedding
model, producing $3{,}072$-dimensional vectors stored in a flat
inner-product FAISS index~\cite{johnson2017faiss}. A single retriever
entry point accepts a query and top-$k$ value and returns records
containing the chunk identifier, lecture and slide indices,
transcript text, slide-image path, and similarity score. Retrievals
are memoised with a SHA-256 disk cache.

Slide-level indexing is central to the system. Each retrieved record
links a textual explanation to a specific slide image, so the same
evidence can ground the narration produced by the engagement module
and the visual content rendered by the audiovisual stack. Coarser
chunking would break this one-to-one link and push the system back
toward text-to-image generation, which our pilot runs identified as a
major source of pedagogical drift.

\subsection{Scaffolding Module}\label{sec:scaffold}

A list of concepts is not sufficient for instruction. A lesson needs
a prerequisite order, a separation between motivation, recall, core
development, application, and synthesis, and a stable structure that
downstream modules can reference. The scaffolding module constructs
this structure explicitly.

The module makes three language-model calls with decreasing
temperatures. The first call, at temperature~0.4, extracts eight to
fifteen concept nodes from retrieved corpus fragments, each with a
short definition and chunk reference. The second call, at
temperature~0.3, predicts directed prerequisite edges with
confidence values in $[0,1]$. The third call, at temperature~0.2,
assigns nodes to the five pedagogical stages:
\emph{motivation}, \emph{prereq-recap}, \emph{core-build},
\emph{application}, and \emph{synthesis}. The decreasing
temperature schedule reflects the decreasing creative freedom across
extraction, linking, and stage assignment.

The predicted prerequisite graph is then post-validated as a directed
acyclic graph. If a cycle is detected, Algorithm~\ref{alg:cyclebreak}
removes the minimum-confidence edge in the cycle until the graph is
acyclic.

\begin{figure}[!t]
\centering
\fbox{\begin{minipage}{0.93\linewidth}
\small
\textbf{Algorithm~1}\enspace \textsc{GreedyCycleBreak}$(E, V)$\\[2pt]
\textbf{Input}: edges $E$ with confidences; concepts $V$.\quad
\textbf{Output}: acyclic edge set $E' \subseteq E$.
\begin{tabbing}
xx\=xxxx\=xxxx\=\kill
1.\> $E' \gets \{\,e \in E : e.\text{from},\, e.\text{to} \in V\,\}$\\
2.\> \textbf{while} $\textsc{FindCycle}(E') \neq \emptyset$ \textbf{do}\\
3.\> \> $C \gets \textsc{FindCycle}(E')$\\
4.\> \> $E_C \gets \{\,e \in E' : (e.\text{from},\, e.\text{to}) \in \text{pairs}(C)\,\}$\\
5.\> \> \textbf{if} $E_C = \emptyset$ \textbf{then return} $E'$\\
6.\> \> $e^\star \gets \arg\min_{e \in E_C}\, e.\text{confidence}$\\
7.\> \> $E' \gets E' \setminus \{e^\star\}$\\
8.\> \textbf{return} $E'$\\
\end{tabbing}
\textsc{FindCycle} is three-colour depth-first search; on a back-edge
the cycle is reconstructed via parent pointers.
\end{minipage}}
\caption{Greedy minimum-confidence cycle break, used after the
prerequisite-linking call to guarantee that the prerequisite graph
is a directed acyclic graph.}
\label{alg:cyclebreak}
\end{figure}

This deterministic repair is a practical compromise. Language models
are effective at proposing prerequisite relations from natural
language, but they are not guaranteed to produce a valid topology.
Removing the weakest edge repairs the graph without another model
call and yields a stable scaffold whose nodes can be referenced by
later modules.

\subsection{Adaptive Style Controller}\label{sec:adapt}

A learner persona combines education level, prior knowledge, urgency,
and motivation. Conditioning the entire generation process on that
persona through a single global prompt collapses these signals into
one undifferentiated tone. The adaptive style controller instead
assigns a per-concept style specification.

Given the concept list and persona, a single language-model call at
temperature~0.3 emits five enumeration-valued fields for each
concept: depth, vocabulary, example density, abstraction, and analogy
use. These fields follow the schema shown in Fig.~\ref{fig:schemas}.
Validation rejects out-of-vocabulary values and replaces them with a
balanced default. The resulting style can vary across concepts within
the same topic; for example, in the filtered-back-projection run, the
\emph{abstraction} field changes from \emph{concrete} for raw
measurements to \emph{mixed} for back-projection and
\emph{abstract} for filtering.

The style specification then conditions the engagement generator
(Section~\ref{sec:engage}) as structured input. It also supports
clean ablation in Section~\ref{sec:experiments}: disabling adaptation
means substituting a default style object, not rewriting the prompt
with an ambiguous instruction such as \emph{be less adaptive}.

\subsection{Engagement Generator}\label{sec:engage}

Effective narration does more than define concepts. It opens with a
reason to pay attention, recalls the relevant prior idea, explains the
new concept at the right level, grounds abstraction with an analogy
when appropriate, and signals the next step. Standard
language-model narrations produce these moves inconsistently. The
engagement generator makes them part of the output contract.

\begin{figure*}[!t]
\centering
\includegraphics[width=0.92\textwidth]{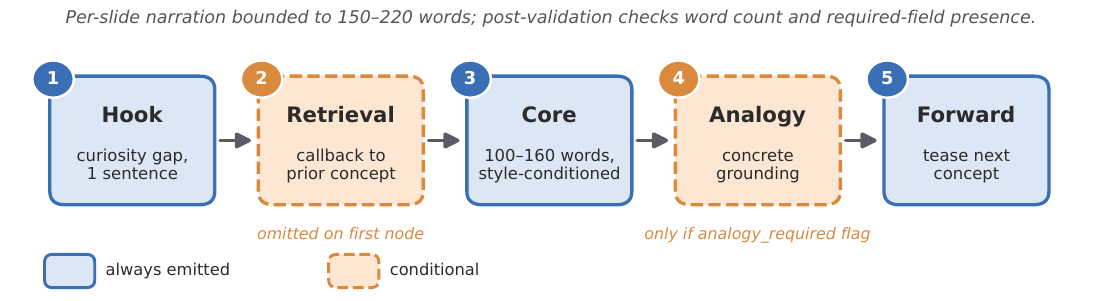}
\caption{Engagement template. Five ordered moves are baked into both
the prompt and the output schema; post-validation rejects short or
malformed narrations. Step~4 is conditional on the
\emph{analogy-required} flag of the style specification; step~2 is
omitted for the first concept.}
\label{fig:engtemplate}
\end{figure*}

For each concept, the module emits a structured narration object that
follows the template in Fig.~\ref{fig:engtemplate}. Each narration is
bounded to 150--220 words and contains, in order, a one-sentence
curiosity gap, a retrieval prompt naming the prior concept
(omitted only for the first node), a 100--160 word core explanation
conditioned on the style specification, an analogy when the style
flag requires one, and a forward hook to the next concept. The object
also stores a visual-hook description for the slide designer and the
corpus chunk identifiers cited by the narration. Post-validation
checks required fields and word count; failures trigger one
regeneration.

Casting engagement as a typed contract makes it measurable. The
objective harness (Section~\ref{sec:harness}) counts analogies and
retrieval prompts with regular expressions over the narration and
cross-checks those counts against the typed fields. The LLM judge
therefore evaluates a consistently structured script rather than
unconstrained prose, reducing avoidable rubric variance.

\subsection{Verifier (Optional)}\label{sec:verify}

Retrieval reduces factual drift but does not eliminate it. The
pipeline therefore includes an optional verifier that can be enabled
at the cost of an additional grounding pass per concept.

The verifier performs three steps. First, an atomic-claim extraction
call at temperature~0.0 converts the narration into a JSON array of
factual claims. Second, a batched verification call compares all $N$
claims against the node's retrieved sources and labels each claim as
\emph{supported}, \emph{contradicted}, or \emph{unsupported}. Third,
a rewrite call at temperature~0.3 removes contradicted and
unsupported claims while preserving the narration structure and
length. Because verification is batched, it requires $O(1)$ calls per
concept rather than $O(N)$ calls. The verifier is disabled in the
reported experiments to control cost and is not on the critical path
of the results.

\subsection{Audiovisual Stack with Corpus-Image Override}\label{sec:av}

The audiovisual stack renders the course blueprint into the final
video and grounds the visual layer through a deterministic
corpus-image override. For each blueprint node $i$, the scaffolding
module records a slide-image hint. If the top-1 retrieved record
exceeds the similarity threshold $\tau{=}0.78$, the hint is the
corresponding chunk identifier; otherwise it is the sentinel string
\emph{generate}. The pipeline passes these hints to the slide-render
stage as a map of the form \{slide identifier $\mapsto$ chunk
identifier\}.

\begin{figure}[!t]
\centering
\includegraphics[width=\linewidth]{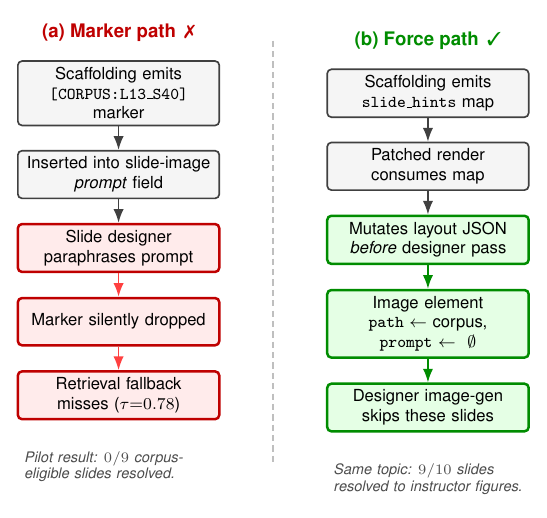}
\caption{Slide-image override. (a)~Marker-based path: a
\texttt{[CORPUS:$\cdot$]} token is inserted into the slide-image
description; the slide designer paraphrases the description and
drops the marker silently. (b)~Force path: the layout JSON is
mutated directly \emph{before} the designer's image-generation pass,
so the corpus image is rendered into the layout unchanged. The
empirical comparison is reported in Section~\ref{sec:results}.}
\label{fig:override}
\end{figure}

A patched render method consumes the map before the slide designer's
image-generation pass executes. For every image element whose slide
has a chunk-identifier hint, the renderer sets the file-path field to
the retrieved corpus image and clears the prompt field
(Fig.~\ref{fig:override}). The text-to-image pass therefore skips
that element, and the original instructor slide is inserted into the
layout unchanged. Nodes marked \emph{generate} use a single-shot
fallback retrieval at the relaxed threshold $\tau'{=}0.65$; if that
also fails, the slide is rendered with a placeholder.

Audio is synthesized with a small text-to-speech model using one
voice across all slides. A forced-alignment pass recovers word-level
timings for each clip. MoviePy then composites the slide images with
their corresponding audio segments and concatenates them into a
single video file.

\subsection{Implementation Details}

The pedagogy modules and LLM judge use Google's fast-tier
general-purpose model in independent sessions so that generation and
evaluation remain auditable. Embeddings are $3{,}072$-dimensional.
All model calls use exponential-backoff retry with five attempts and
a two-second base delay. To keep wall-clock time tractable, the
slide-designer review threshold is relaxed from $9.0$ to $7.5$, and
the maximum number of review rounds is reduced from $5$ to $2$.
Manual inspection found no visible degradation in slide quality under
this setting.

\section{The BMED 2300 Benchmark Corpus}\label{sec:corpus}

To evaluate course-grounded pedagogical video generation, we package
Dr.~Wang's BMED~2300 Bio-Imaging course as a structured benchmark
artefact. Table~\ref{tab:corpus} summarizes the corpus and retrieval
index parameters.

The chunker walks each lecture directory and emits one record per
\emph{(slide, narration)} pair. Each record is assigned a stable
identifier of the form \textsc{L}\textit{N}\textsc{\_S}\textit{M},
and case-insensitive image matching covers JPG, jpg, jpeg, and png
files. Slide-granular chunking is the key design choice: each
embedded vector preserves a one-to-one link between a textual
explanation and its corresponding slide image. This allows retrieval
to ground both the narration and the visual layer from the same
record. The corpus and FAISS index will be released with the
chunking script under the same license as the released starter kit.

\begin{table}[!t]
\centering
\caption{BMED~2300 corpus and retrieval index parameters.}
\label{tab:corpus}
\renewcommand{\arraystretch}{1.1}
\begin{tabular}{lr}
\toprule
Lectures                           & 23 \\
Slides (chunks)                    & $1{,}116$ \\
Mean transcript words per slide    & $\sim$80 \\
Embedding dimension                & $3{,}072$ \\
Index type                         & flat inner-product (FAISS) \\
Top-$k$ at retrieval               & 8 \\
Image-override threshold $\tau$    & 0.78 \\
Image-fallback threshold $\tau'$   & 0.65 \\
\bottomrule
\end{tabular}
\end{table}

\section{Evaluation Methodology}\label{sec:harness}

The evaluation harness combines rubric-based LLM judging with
regex-grounded objective metrics. This pairing allows subjective
pedagogical scores to be checked against measurable properties of
the generated artefact.

\textit{LLM-judge rubric.}
Three independent rubric prompts evaluate \emph{scaffolding},
\emph{adaptive teaching}, and \emph{engagement} on a 1--5 scale with
explicit anchors at scores 1, 3, and~5. The scaffolding rubric
receives the concept tree and narrations, the adaptive rubric
receives the persona and narrations, and the engagement rubric
receives the narrations and visual hooks. Each rubric is scored
$n{=}3$ times, and we report the median together with all per-rep raw
scores for transparency.

\textit{Objective metrics.}
For each run, we compute the number of concepts, prerequisite-graph
depth, Flesch reading ease~\cite{flesch1948reading}, words per
minute from forced-alignment timings, visual element density per
slide, total narration word count, and counts of analogies, retrieval
prompts, and questions. The engagement counts are obtained by regular
expression over the concatenated narration. The analogy pattern
matches \emph{like}, \emph{as if}, \emph{imagine}, \emph{think of},
\emph{analogous}, \emph{just as}, and \emph{similar to}. The
retrieval-prompt pattern matches \emph{recall}, \emph{earlier we},
\emph{as we saw}, \emph{remember when}, and \emph{back in}. Regex
counts are validated on a sample against the typed analogy and
retrieval-prompt fields of the enriched script.

\textit{Inter-rep agreement (\emph{within-judge}).}
We measure the reproducibility of a single Gemini-Flash judge across
$n{=}3$ repetitions. This analysis characterizes \emph{within}-judge
stability only; agreement across model families is treated as a
separate limitation and future validation step in
Section~\ref{sec:disc}. Across the thirty cells in the main ablation
(five topics, three PCK dimensions, and two variants), twenty-six
cells have identical scores across all repetitions, and four cells
differ by exactly one point on a single repetition. The mean per-cell
standard deviation is $0.063$, and Krippendorff's
$\alpha$~\cite{krippendorff2011alpha}, treating repetitions as
raters, is approximately~$0.97$. Thus, within-judge variance is
negligible at this rubric resolution, while cross-judge-family
agreement remains the next required check.

\section{Experimental Setup}\label{sec:experiments}

\textit{Topics and persona.}
We evaluate on five in-corpus BMED~2300 topics: filtered back
projection, T1/T2 relaxation in MRI, ultrasound transducers and beam
formation, X-ray photon attenuation and contrast, and CT
image-reconstruction artifacts. All runs use one freeform learner
persona: \emph{Education level: undergraduate biomedical engineering.
Prior: linear algebra, introductory physics. Urgency: standard.}
Using a single persona isolates the ablation effect from
per-persona variation, but it also limits generality, as discussed in
Section~\ref{sec:disc}.

\textit{Systems compared.}
\textbf{Ours~(full)} enables scaffolding, adaptive style, engagement,
and the slide-image override, with the verifier disabled.
\textbf{No~Engagement} keeps the same scaffold, adaptive style
assignment, and slide-image override, but disables the engagement
module. In this variant, each node receives a single-sentence
fallback narration derived from the concept definition. The
comparison therefore isolates the effect of the engagement contract
while holding the conceptual scaffold and adaptive style assignment
fixed.

\textit{Evaluation protocol.}
Each variant is run once on each topic, producing five videos per
variant and ten videos in total. Each video is scored by the
LLM-judge harness with $n{=}3$ repetitions and by the deterministic
objective harness. We report aggregate results as variant-level
means.

\section{Results}\label{sec:results}

\subsection{Variant means and per-topic detail}

\begin{figure*}[!t]
\centering
\includegraphics[width=0.92\textwidth]{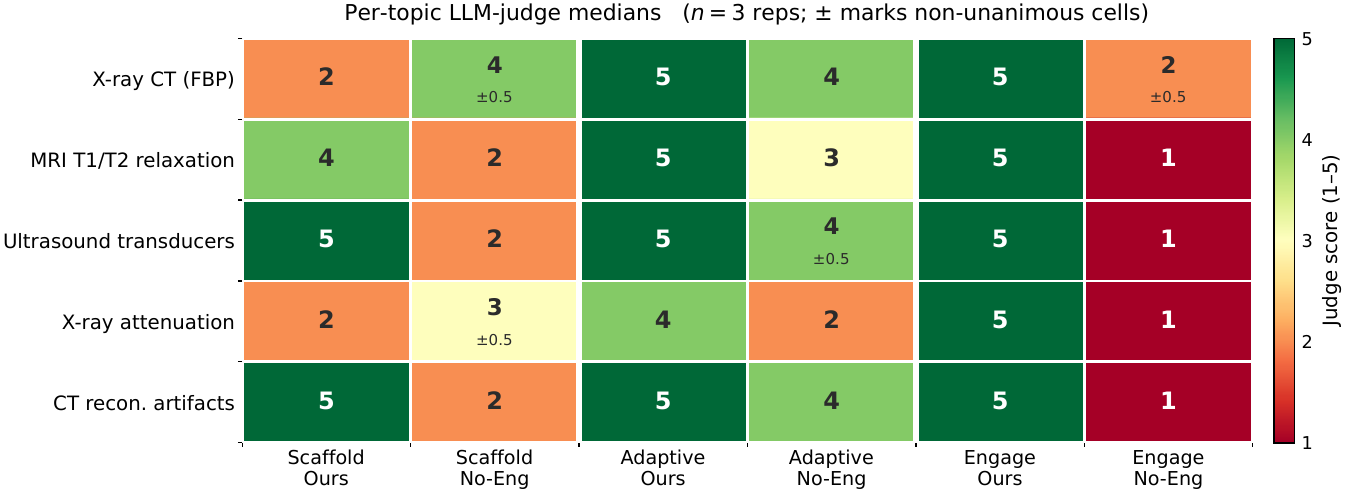}
\caption{Per-topic LLM-judge medians for both variants. Cells show
the median rep ($\pm$ inter-rep standard deviation). Engagement is
essentially binary across topics (5 versus 1--2); the adaptive
dimension shows a consistent 1--2 point gap. Scaffolding has the
most cross-topic variance, indicating that prerequisite extraction
quality is topic-sensitive.}
\label{fig:heatmap}
\end{figure*}

Table~\ref{tab:means} summarizes the aggregate results over five
topics, and Fig.~\ref{fig:heatmap} shows the corresponding per-topic
LLM-judge medians. The improvements are consistent across topics,
rather than being driven by a single outlier.

\begin{table}[!t]
\centering
\caption{Variant means over five topics ($n{=}5$). Judge scores are 1--5
medians of three reps with inter-rep (\emph{within}-judge) standard
deviation in parentheses; objective metrics are arithmetic means.
Only the engagement module is ablated here; scaffolding-off and
adaptive-off ablations, multi-persona runs, and a between-judge
agreement probe are planned follow-ups (Section~\ref{sec:disc}).}
\label{tab:means}
\renewcommand{\arraystretch}{1.1}
\setlength{\tabcolsep}{3pt}
\footnotesize
\begin{tabular}{lcccccc}
\toprule
& \multicolumn{3}{c}{LLM judge (median, $\sigma$)} & \multicolumn{3}{c}{Objective} \\
\cmidrule(lr){2-4}\cmidrule(lr){5-7}
Variant & Scaff. & Adapt. & Eng. & Analog. & Retr. & Flesch \\
\midrule
Ours (full) & 3.60 (0.0) & \textbf{4.80} (0.0) & \textbf{5.00} (0.0)
            & 20.0 & 18.6 & 38.0 \\
No Eng. & 2.60 (0.2) & 3.40 (0.1) & 1.20 (0.2)
        & \phantom{0}0.2 & \phantom{0}0.0 & 19.8 \\
$\Delta$ & $+1.00$ & $+1.40$ & $+3.80$ & $+19.8$ & $+18.6$ & $+18.2$ \\
\bottomrule
\end{tabular}
\end{table}

\begin{figure*}[!t]
\centering
\includegraphics[width=0.92\textwidth]{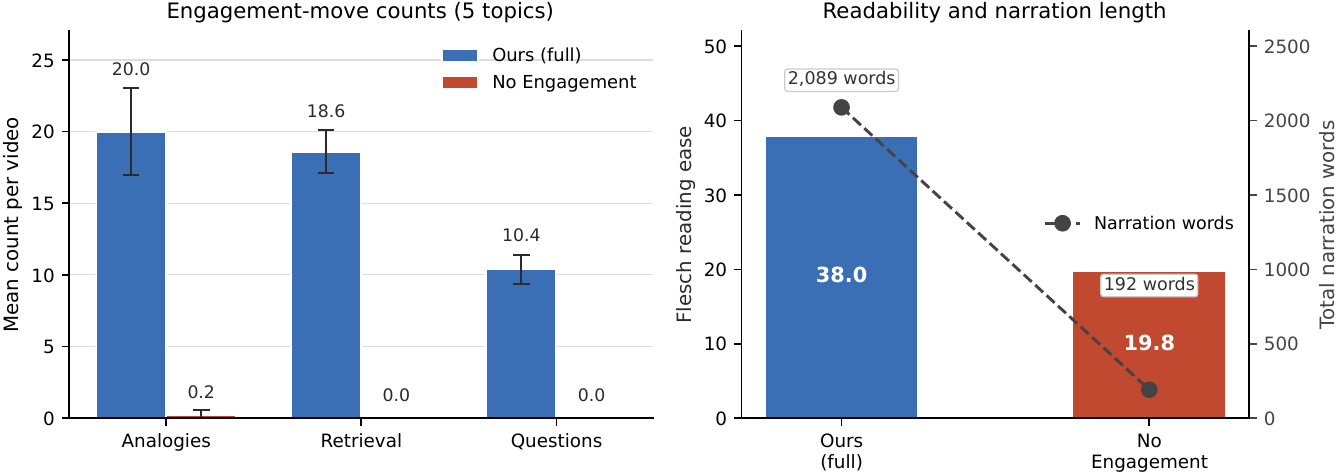}
\caption{Mean engagement-move counts per video over the five topics
(left) and mean Flesch reading ease per variant with total narration
words on a secondary axis (right). The full pipeline produces an
order-of-magnitude longer narration that is also more readable at
the undergraduate level.}
\label{fig:moves}
\end{figure*}

\textit{Engagement is the decisive contributor.}
Disabling the engagement module reduces the engagement score from
5.00 to 1.20. The objective metrics show the same pattern:
analogies fall from 20.0 to 0.2 per video, and retrieval prompts fall
from 18.6 to 0.0 per video (Fig.~\ref{fig:moves}). Thus, the judge is
not merely rewarding longer text; the structured engagement moves
that define the contract are present in the full system and absent in
the ablation.

\textit{Engagement also realizes adaptation.}
The adaptive score drops from 4.80 to 3.40 when engagement is
disabled. This indicates that the adaptive style controller alone is
not sufficient: it specifies persona-conditioned depth, vocabulary,
abstraction, and example density, but the engagement generator is the
module that realizes those choices in the narration. This interaction
explains why engagement affects both the engagement and adaptive
dimensions.

\textit{Readability improves with structured narration.}
Removing engagement lowers mean Flesch reading ease from 38.0 to
19.8. The fallback narrations are short concept-definition fragments,
with average narration length falling from 2089 words to 192 words.
The full pipeline is therefore more expensive to render, but it
produces a substantially richer and more readable instructional
artifact.

\subsection{Concept graph inspection}

\begin{figure*}[!t]
\centering
\includegraphics[width=0.92\textwidth]{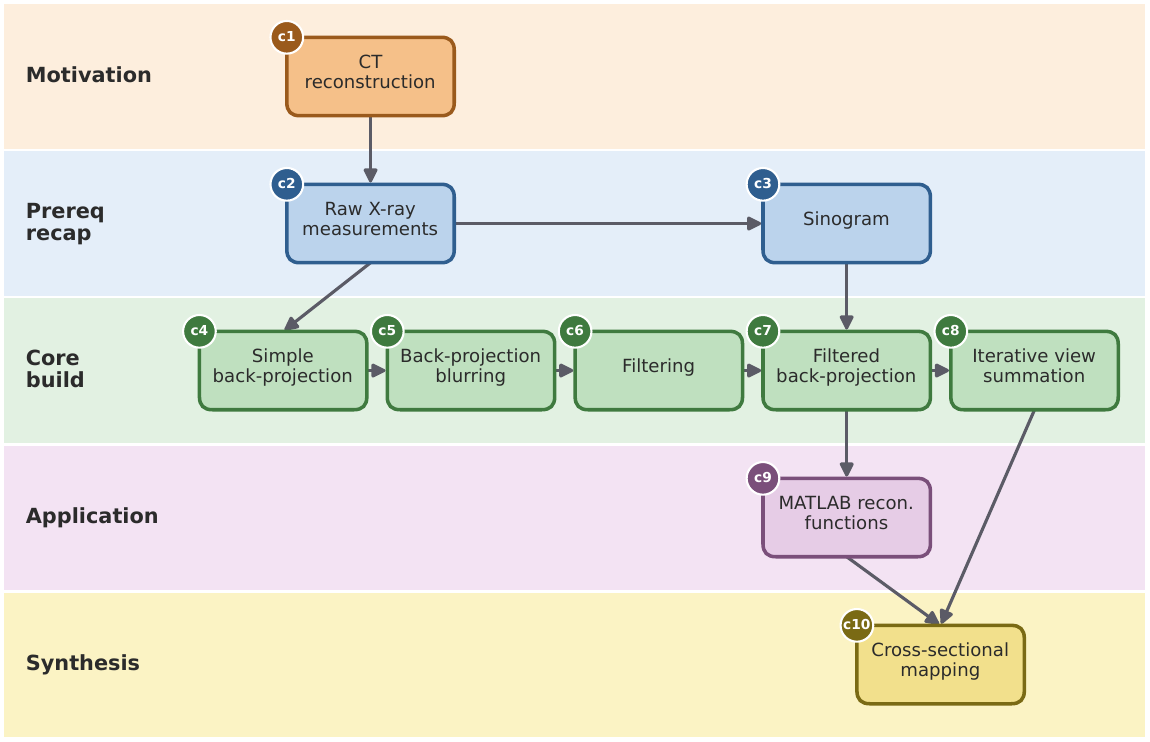}
\caption{Concept graph produced by the scaffolding module for the
filtered-back-projection topic. Nodes are coloured by stage. The
cycle-break algorithm has resolved the prerequisite graph;
cross-stage edges form the build chain (simple back-projection
$\to$ blurring $\to$ filtering $\to$ filtered back-projection).}
\label{fig:dag}
\end{figure*}

The typed pipeline also produces auditable intermediate artifacts.
Fig.~\ref{fig:dag} visualizes the concept graph generated by the
scaffolding module for filtered back projection. The
\emph{prereq-recap} stage contains the early dependencies, including
raw measurements and the sinogram. The \emph{core-build} stage then
organizes the back-projection, blurring, and filtering sequence, and
an \emph{application} node closes the lesson. Although the
cycle-break procedure runs for every topic, none of the five
evaluation topics required edge removal. Cycles appeared only in
pilot runs on out-of-corpus topics, suggesting that the algorithm is
primarily a safety mechanism rather than a routine repair step.

\subsection{Qualitative narration comparison}

\begin{figure*}[!t]
\centering
\includegraphics[width=0.92\textwidth]{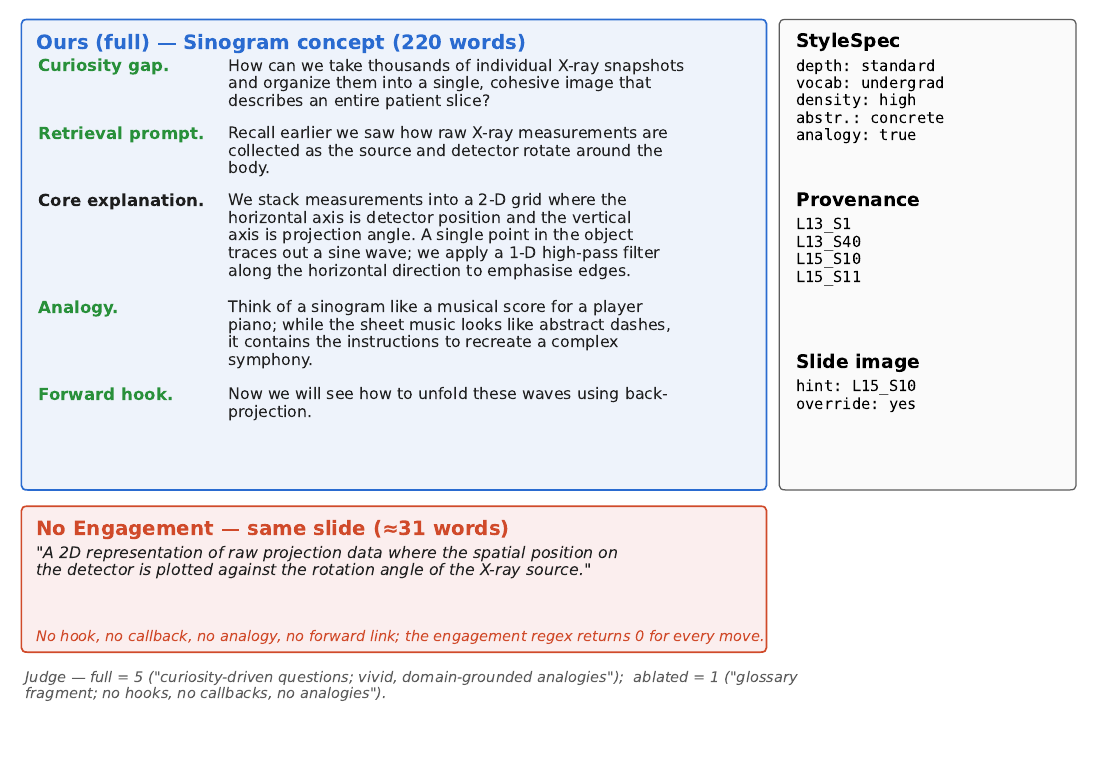}
\caption{Qualitative comparison on the sinogram concept of the
filtered-back-projection topic. Top: the full pipeline emits a
220-word, structurally complete narration with the five engagement
moves explicitly tagged. Bottom: the ablated variant emits a
$\sim$31-word glossary fragment. Right column: the typed style
specification and the provenance chunk identifiers consumed by the
engagement stage.}
\label{fig:qual}
\end{figure*}

Fig.~\ref{fig:qual} compares the same sinogram slide under the full
pipeline and the No-Engagement ablation. The difference is structural,
not merely stylistic. The full pipeline produces a narration with a
hook, prior-concept callback, analogy, and forward link. The ablated
variant produces a short glossary-like fragment with none of these
moves. This is exactly the contrast captured by the engagement-move
regular expressions and by the judge reasoning.

\subsection{Slide-image override}

The deterministic slide-image override provides a separate systems
result that is not reflected in the aggregate scores. As described in
Section~\ref{sec:av}, the marker-based approach failed to preserve
corpus-slide references after the slide designer paraphrased the
image prompt. On the filtered-back-projection topic, that path
resolved $0$ of $9$ corpus-eligible slides. Replacing it with the
slide-hints map resolved $9$ of $10$ slides on the same topic. We
therefore report the override as a deterministic grounding property:
it changes slide reuse from a fragile prompt convention into an
explicit rendering decision.

\section{Discussion}\label{sec:disc}

The main result is that the engagement contract has effects beyond
engagement itself. A module designed to introduce hooks, callbacks,
analogies, and forward links also improves persona fit and
readability. This suggests that, in pedagogical video generation,
adaptive style and engagement should not be treated as independent
surface features. The adaptive controller specifies the desired
instructional style, but the engagement generator is where that style
is realized in learner-facing language.

This finding has two implications. First, engagement should be
implemented as an explicit typed contract rather than as a global
instruction to ``be more engaging.'' Without a structured output
schema, the model may produce fluent narration while omitting the
instructional moves that sustain attention and connect concepts.
Second, PCK dimensions should be evaluated jointly as well as
individually. Scaffolding, adaptation, and engagement interact in the
final artifact, and scoring them in isolation can underestimate their
dependence on one another.

\section{Limitations and Future Work}\label{sec:limits}

The study has several limitations. First, the judge is drawn from the
same model family as the generator, which raises self-preference
risk~\cite{panickssery2024selfpref}. Although the measured
\emph{within}-judge variance is small
($\bar{\sigma}{=}0.063$, $\alpha{\approx}0.97$), this does not
establish agreement across model families. The most important next
step is therefore a \emph{between}-judge evaluation using additional
judges from different model families~\cite{kim2024prometheus2,
liu2023geval}, with agreement and per-judge medians reported across
all ten generated videos.

Second, the current evaluation does not include human learners. This
limits the strength of any claim about learning effectiveness. A
planned within-subjects study with $n{=}20$ undergraduate
biomedical-engineering students, Latin-square topic ordering, and
5-point Likert ratings for each PCK dimension is the next step before
making broader claims about educational impact.

Third, the experimental scope is narrow. The reported study uses five
in-corpus topics and one learner persona. Out-of-corpus topics fall
back to placeholder slides and therefore lose the main benefit of
course-grounded visual reuse (Section~\ref{sec:av}). The typed style
specification already supports multi-persona evaluation, but fresh
runs are required to test robustness across learner profiles and
topic types.

Fourth, the present ablation isolates only the engagement module.
Scaffolding-off and adaptive-off conditions remain future work, so
the current results should be interpreted as a focused contrast
between the full system and a no-engagement variant, not as a full
causal decomposition of all modules.

Fifth, the optional verifier is disabled in the reported experiments
to control cost. As a result, factual correctness is supported by
retrieval but not independently verified. Because the verifier uses a
batched design with a constant number of calls per concept, a
verifier-on ablation is a tractable follow-up.

Finally, the released starter kit's cursor module depends on
GPU-resident vision-language models and is omitted on our development
host. The generated videos therefore use static slides with
synchronized audio. The slide-image override improves instructor
authenticity, but it can reduce visual diversity; an
image-generation fallback for out-of-corpus topics is left for a
future version.

\section{Reproducibility}\label{sec:repro}

All code, the pre-built FAISS index, typed-schema definitions,
LLM-judge rubrics, regex-based objective metrics, and five-topic
evaluation outputs will be released under the same license as the
released starter kit upon publication. The released outputs include
10~videos, per-run intermediate artifacts, and the aggregated CSV.

The BMED~2300 corpus, including transcripts and slide images, is
released with the chunking script and embedding configuration so that
the FAISS index can be rebuilt deterministically. All pedagogy
modules expose Boolean enable flags controlled by a single YAML
configuration, allowing the reported ablations to be reproduced
end-to-end by changing one field. Generation and evaluation use
independent Gemini-Flash sessions with fixed temperatures and a
five-attempt exponential-backoff retry policy. Embeddings are
$3{,}072$-dimensional, and the retrieval cache is keyed by the
SHA-256 hash of the query.

\section{Conclusion}\label{sec:conc}

This paper argues that pedagogy must be treated as a first-class
constraint in generative video systems. We introduced a
retrieval-augmented pipeline that replaces prompt chaining with typed
intermediate representations and reuses corpus slides whenever
retrieval confidence is high. The resulting system makes scaffolding,
adaptation, engagement, and visual grounding inspectable and
ablation-friendly.

On five biomedical-imaging topics, removing the engagement contract
reduces the LLM-judge engagement score from 5.00 to 1.20, the
adaptive score from 4.80 to 3.40, and Flesch readability from 38 to
20. Separately, the deterministic slide-image override converts a
$0/9$ marker-based corpus-grounding failure into a $9/10$
corpus-image success on the same topic. These results indicate that
structured engagement is the load-bearing PCK component in this
setting, and that typed instructional contracts make this effect
measurable. The remaining gaps---between-judge robustness, human
learner evaluation, broader persona coverage, additional module
ablations, and an image-generation fallback for out-of-corpus
topics---are direct extensions of the current architecture.

\appendices

\section{Engagement rubric anchors}

\textit{Score 5}: every slide opens with an explicit hook, such as a
question, paradox, or curiosity gap; at least eighty percent of slides
recall a prior concept by name; and analogies appear when the
abstraction level warrants them. \textit{Score 3}: roughly half of
the slides include a hook, callbacks are sporadic, and one or two
analogies appear. \textit{Score 1}: the narration contains no hooks,
callbacks, or analogies and reads as a definition list. The
scaffolding and adaptive rubrics are anchored analogously.

\section{Per-topic LLM-judge scores}

Table~\ref{tab:pertopic} reports per-topic median scores for
scaffolding, adaptive teaching, and engagement under the two
variants. Engagement is essentially binary across topics, with the
full system scoring 5 and the no-engagement variant scoring 1--2.
Scaffolding shows the largest cross-topic variation.

\begin{table}[h]
\centering
\caption{Per-topic LLM-judge medians: scaffolding / adaptive / engagement.}
\label{tab:pertopic}
\renewcommand{\arraystretch}{1.15}
\begin{tabular}{l l c c}
\toprule
ID  & Topic                          & Full        & No Eng. \\
\midrule
t01 & X-ray CT --- FBP              & 2 / 5 / 5   & 4 / 4 / 2 \\
t02 & MRI T1/T2 relaxation          & 4 / 5 / 5   & 2 / 3 / 1 \\
t03 & Ultrasound transducers        & 5 / 5 / 5   & 2 / 4 / 1 \\
t04 & X-ray photon attenuation      & 2 / 4 / 5   & 3 / 2 / 1 \\
t05 & CT reconstruction artifacts   & 5 / 5 / 5   & 2 / 4 / 1 \\
\bottomrule
\end{tabular}
\end{table}

\section{Engagement-move counts per topic}

Table~\ref{tab:moves} reports per-topic regex counts of analogies,
retrieval prompts, and questions for the two variants. The full
system consistently produces all three engagement-move types, while
the no-engagement variant reduces them to near zero.

\begin{table}[h]
\centering
\caption{Engagement-move counts per topic and variant.}
\label{tab:moves}
\renewcommand{\arraystretch}{1.15}
\setlength{\tabcolsep}{3pt}
\begin{tabular}{l ccc ccc}
\toprule
& \multicolumn{3}{c}{Ours (full)} & \multicolumn{3}{c}{No Engagement} \\
\cmidrule(lr){2-4}\cmidrule(lr){5-7}
Topic  & Analog. & Retr. & ? & Analog. & Retr. & ? \\
\midrule
t01   & 20 & 19 & 10 & 1 & 0 & 0 \\
t02   & 15 & 20 & 12 & 0 & 0 & 0 \\
t03   & 22 & 18 & 10 & 0 & 0 & 0 \\
t04   & 19 & 20 & 11 & 0 & 0 & 0 \\
t05   & 24 & 16 & \phantom{0}9 & 0 & 0 & 0 \\
\midrule
mean  & 20.0 & 18.6 & 10.4 & 0.2 & 0.0 & 0.0 \\
\bottomrule
\end{tabular}
\end{table}

\section{Sample provenance trace}

For the sinogram concept of the filtered-back-projection topic, the
engagement generator returned the provenance list \textsc{L13\_S1},
\textsc{L13\_S40}, \textsc{L15\_S10}, and \textsc{L15\_S11}. The
narration cites the sinogram structure (\textsc{L13\_S1}), the
back-projection lecture (\textsc{L15\_S10}), and the
filtered-back-projection slide (\textsc{L15\_S11}). These retrieved
records were also used by the slide-image override at threshold
$\tau{=}0.78$ to populate the slide with the corresponding instructor
figures.

\bibliographystyle{IEEEtran}
\bibliography{refs}

\end{document}